\documentclass[fleqn,usenatbib]{mnras}

\usepackage{newtxtext,newtxmath} 
\usepackage[T1]{fontenc}
\DeclareRobustCommand{\VAN}[3]{#2}
\let\VANthebibliography\thebibliography
\def\thebibliography{\DeclareRobustCommand{\VAN}[3]{##3}\VANthebibliography}

\usepackage{graphicx}	

\usepackage{epstopdf}

\usepackage[normalem]{ulem}

\title{Oscillations in Gas-grain Astrochemical  Kinetics} 
\author[Dufour, Charnley \& Lindberg]{Gwena\"elle  Dufour,$^{1,2,3}$
Steven B. Charnley$^{1}$
and Johan E. Lindberg$^{4,1}$
\\
$^{1}$Astrochemistry Laboratory, Code 691, NASA Goddard Space Flight Center, Greenbelt, MD 20771, USA\\
$^{2}$Department of Physics, Catholic University of America, Washington, DC 20064, USA\\
$^{3}$Leiden Observatory, Leiden University, Leiden, The Netherlands\\
$^{4}$NASA Postdoctoral Program Fellow at NASA Goddard Space Flight Center, administered by Universities Space Research Association through a contract with NASA.
}

\date{Accepted 2022 December 19. Received 2022 December 15; in original form 2022 February. Published 17 January 2023.}
\pubyear{2023}

\begin{document}
\label{firstpage}
\pagerange{\pageref{firstpage}--\pageref{lastpage}}
\maketitle


\begin{abstract} 
  
We have studied gas-grain chemical models of interstellar clouds to search for nonlinear dynamical evolution. 
A prescription is given for producing oscillatory solutions when a bistable solution exists in the gas-phase chemistry  and we demonstrate the existence of limit cycle and relaxation oscillation solutions.   
As the autocatalytic chemical processes underlying these solutions are common to all models of interstellar chemistry, the occurrence of these solutions should be widespread.  We briefly discuss the implications for interpreting molecular cloud composition with time-dependent models and some future directions for this approach.  
\end{abstract} 



\section{Introduction} \label{sec:intro}

The  systems of differential equations governing the kinetics of chemical reaction networks  can exhibit a wide range of nonlinear dynamical phenomena: limit cycles, complex oscillations and chaos can all emerge from stable stationary solutions, subject to variation of externally-controlled bifurcation parameters (\citet{1996OxPres...314...G,1991Oxford..S}).
One signpost of  possible non-linear chemical kinetics is the presence of bistability in the stationary, steady-state, solutions (i.e. the  fixed points) of the  associated system of differential equations. In chemical systems, multistability can arise when autocatalysis is present, i.e. when one species, the autocatalyst, can  take part in reaction sequences which promote its formation. When  the stationary solutions become unstable a Hopf bifurcation emerges and  limit cycle oscillations can occur.
 
 In the  simplest, purely gas-phase, chemical models of dense interstellar clouds,  the chemistry is driven by cosmic-ray ionization of H$_2$ and evolves to approach a steady-state  on time-scales of typically $\sim \rm few \times 10^6$ years. 
   In all cases the chemistry tends to follow well-defined evolutionary pathways and, when making comparisons with observed molecular abundances,  it has become common to label this as `late-time' chemistry. By contrast, `early-time' chemistry occurs around $\sim \rm few \times 10^5$ years (\citet{1984ApJS...56..231L}) and is strongly dependent on the initial conditions. 
 The accretion and desorption of atomic and molecular material  onto and from dust grains  leads to  the formation of molecular ice mantles. 
When desorption is inefficient,  species can be completely lost from the gas (e.g. \citet{leung1984synthesis}); when desorption mechanisms can compete with accretion a quasi-steady-state can  persist (e.g. \citet{1993MNRAS.260..635W,1998MNRAS.298..562W,2001A&A...378.1024C}). 
This  occurs at a time-scale determined by the physical conditions such as the gas density, the grain-size distribution and the dust physics, but the quasi-steady-state is typically comparable to, or longer than, `early-time' chemistry  and shorter than `late-time' chemistry.
Hence, the  possibility of oscillatory and chaotic solutions manifesting themselves would have major implications for modeling, interstellar clouds and  astrochemistry in general.

The existence of  bistable solutions in interstellar gas-phase chemistry  has been recognized for some time (\citet{1992MNRAS.258P..45P,1993ApJ...416L..87L}) and has been the subject of several studies (e.g. \citet{1995A&A...297..251L,1995A&A...302..870L,1998A&A...334.1047L,2000RSPTA.358.2549P,2001A&A...370..557V,2003A&A...399..583C,2006A&A...459..813W}). This raises the prospect that oscillations and chaos could be common but heretofore unrealized solutions to  well-studied chemical models.  
 However,  it is only recently that limit cycle oscillations have been reported.\footnote{\citet{1993ApJ...416L..87L}  did previously report one case of  limit cycle oscillations   in a gas-phase dense cloud model.}
\citet{2020A&A...643A.121R} showed that sustained chemical oscillations,  could occur in purely gas-phase models of dense molecular clouds. They found that the oscillations originated in the nitrogen chemistry for temperatures in the range $T \approx 7-15$K and was  controlled by the H$_2$ ortho/para ratio. Roueff \& Le Bourlot presented an overview of the important chemical reactions in their model, identifying deuterium chemistry as   an important component, but no underlying  mechanism, such as emergence from a bistable state, was identified.   

 \citet{2019ApJ...887...67D}  (Paper I)  demonstrated that the known  astrochemical bistable solutions  have an autocatalytic origin in oxygen chemistry. \citet{2021ApJ...909..171D} (Paper I \& II) further  showed that autocatalysis and bistability can also occur in the nitrogen and carbon chemistry  of dense molecular clouds. 
 In each case,  autocatalysis involves  dimer-autocatalyst pairs: O$_2$-O, N$_2$-N, H-C$_2$H$_2$, with other autocatalytic cycles also being possible in carbon chemistry (e.g. C$_2$-C). It is known that  limit cycle oscillations can be induced in a bistable chemical system in which species crucial to autocatalysis are 
removed and re-introduced by an additional  feedback process which occurs on a time-scale longer or comparable to that of the chemical reactions, i.e. the feedback renders the stationary solutions unstable  (e.g. \citet{2003DaltTra...7..1201S}).
 In interstellar clouds, the exchange of atoms and molecules between gas and dust grains naturally provides this process. 
    
The plan of this article is as follows. In  section \ref{sec:theory} we summarize the mathematical structure of the model with respect to bistable gas-phase solutions and the gas-grain interaction.  The parameters adopted in a reference chemical model are presented in section \ref{sec:model}. In section \ref{sec:heho} we show that introducing accretion and desorption of atomic and molecular oxygen into a simple reduced model of interstellar chemistry, with known autocatalytic pathways and bistable solutions, leads to chemical oscillations.   We then show that  oscillatory solutions are also present in more realistic models of dense cloud chemical evolution (section \ref{sec:chemosc}). Extension of the modeling are briefly discussed  in section \ref{sec:discuss} and conclusions are given in section \ref{sec:conc}.  
   
\section{Theory}\label{sec:theory}

\subsection{Bistable Solutions in Gas-phase Chemistry} \label{sec:bistable}  
  
The chemical evolution of $N$ gas-phase chemical species in static dense molecular gas is obtained by solving the system of  autonomous nonlinear ordinary  differential equations (ODEs)

\begin{multline}
 {\dot y _i} ~ = ~ G_i ~ = ~
 n_{\rm H} \Bigl \lbrack  \sum _j \sum _m k_{jm} y_j y_m ~-~ y_i \sum _s k_{is} y_s
\Bigr \rbrack  \\
~+~ \sum _u \beta _u  y_u ~-~y_i \sum _w \beta _w ; ~~~~~~~i=1,...,N
\end{multline} 
where $n_{\rm H}$ is the density of hydrogen nuclei ($n_{\rm H} (=n(\rm H)+2n(H_2)$; fractional abundance of species $i$,  $y_i(t)$,  is equal to $n_i/n_{\rm H}$, for number density  $n_i$; and  $k_{jm}$, $\beta _u$, $\beta _w $ are
reaction rate coefficients for various bimolecular and unimolecular processes. 
     This system is solved subject to prescribed values of the cosmic ray ionization rate
  $\zeta$, the hydrogen nucleon number density $n_{\rm H}$, the gas temperature $T$, the visual extinction, $A_V$, 
  and the total elemental abundances, $Y^{\rm T}_j$, of the major volatile elements and the refractory metals (e.g. $j$=C,O,N,S,Si,Na),
  The elemental abundances are defined  relative to a reference value, values of $X_j$  (e.g. a diffuse cloud line of sight) and a depletion factor, $\delta_j$, as $Y^{\rm T}_j=X_j \delta_j$;  we used the same values of $X_j$  as in papers I and II. 
 The reaction rate coefficients are assumed to remain constant in time.  For a defined set of initial conditions on the $N$ species,  $y_i(0)$,  the time-dependent solution can be obtained with a stiff ODE solver such as DVODE (\citet{2005Ap&SS.299....1N}).

It is necessary to examine the  steady-state solutions for bistability and other bifurcation phenomena. The system of differential equation can only yield stable solutions and so,  as both the stable and unstable solutions are necessary to fully characterize the  bifurcations, we require to use Newton-Raphson iteration to solve the system   
        
\begin{multline}
 {{ G} _i}({\bf y}; \zeta/n_{\rm H},  \zeta_{\rm He}, T, A_V, \beta_{\rm crp}, Y^{\rm T}_{C}, Y^{\rm T}_{O},Y^{\rm T}_{N},
 Y^{\rm T}_{S},  Y^{\rm T}_{Na}  )  = 0~~~~~~\\i=1,...,N.
 \label{NR}
\end{multline}

  The appearance/disappearance of bistable solutions is principally controlled by variation of the bifurcation parameters:   
    the cosmic-ray ionization of H$_2$ and He,  through the ratios  $\zeta/n_ {\rm H}$ and $\zeta_{\rm He}/n_ {\rm H}$, the presence of cosmic ray-induced photons $\beta_{\rm crp}$,  the relative  elemental depletions $Y^{\rm T}_j$,   and  the H$_3^+$ electron recombination rate, $\rm \alpha_3$  (\citet{1993ApJ...416L..87L, 1995A&A...297..251L, 1998A&A...334.1047L, 2000RSPTA.358.2549P,2006A&A...459..813W,2019ApJ...887...67D,2021ApJ...909..171D}).

\subsection{Gas-grain interaction in cold interstellar clouds}\label{sec:gasgrain}   
      
     For each neutral species $i$, the coupled evolution of gas and grain-surface abundances  is described by  the 
differential equations 

\begin{equation}
 {\dot y _i} = { G _i} - \lambda_i  y_i + \xi _i  g_i
 \label{eqnydot}
\end{equation}
\begin{equation}
 {\dot g _i} = \lambda_i  y_i -  \xi _i  g_i
 \label{eqngdot}
\end{equation}
where $g_i$ is the fractional abundance of  $i$ currently resident on dust. 
Sticking collisions of gaseous species with dust occur at the accretion rate, $\lambda _i$,
and surface species leave the dust grain at the desorption rate,   $\xi _i$   (e.g. \citet{1990MNRAS.244..432B}).
These differential equations satisfactorily account for the exchange of material between the gas and dust and, as we show,  are sufficient to investigate the effects of dust on bistable solutions without attempting to explicitly consider grain-surface chemical reactions (see $\S$ \ref{sec:gsreac} ).  
The effect of dust on bifurcations is that, on time-scales longer than the gas particle accretion accretion time ($\sim 1/\lambda_i $), the elemental abundances in the gas,  $Y_j$,  are set by the desorption rate of particles containing element $j$, and so we can expect   $Y_j < Y^{\rm T}_j$.
 Thus, for each pair of species involved in an autocatalytic cycle,  $X$ and $X_2$, the ratios $\lambda_{\rm X} / \xi_{\rm X}$ and $\lambda_{\rm X_2} / \xi_{\rm X_2}$ will be bifurcation parameters. 
We model the  gas-grain interaction as follows.

\subsubsection{Accretion}  \label{sec:acc}

Neutral atoms and molecules collide and stick to a distribution of  dust grains  at an accretion rate given  by 
 
\begin{equation}
\lambda_i =   ({8 k T \over \pi M_{i} m_{\rm H}})^{1/2} S_i <n_{\rm gr}(a)\sigma(a)>
\label{eqn:acc1}
\end{equation}
where $k$ is the Boltzmann constant,  ${M_{i}}$ is the molecular weight  (in a.m.u), $S_i$ is the sticking efficiency, 
$n_{\rm gr}(a)$ is the  number  density  of grains of radius $a$, and $\sigma(a) $ the grain collision cross-section. 
 Assuming a single-size distribution of spherical grains with $a = 0.1$\micron, $n_{\rm gr}(a) = 10^{-12}n_ {\rm H}$,  and $S_i =1$ for neutral species heavier than He, one has 

\begin{equation}
\lambda_i = 1.45 \times 10^{4}  D_{\rm gr} ( { T \over {M_{i}} } )^{1/2}  n_ {\rm H}
\label{eqn:acc2}
\end{equation}
where $D_{\rm gr} $ is the total available dust cross-section  
 
\begin{equation}
D_{\rm gr}  =  { <n_{\rm gr}(a)\sigma(a)> \over  n_ {\rm H} }  
\label{eqn:dgr}
\end{equation}
    
 In dense clouds, electron sticking leads to negatively-charged grains dominating  the dust distribution  (\citet{1990MNRAS.243..103U}).  We assume that when atomic ions collide with grains they are neutralized and released into the gas. Molecular ions will not be affected by grain interaction as electron recombination is more efficient  at the densities of interest. As shown in Paper I, only grain neutralization of $\rm H^+$ and $\rm S^+$ can affect bistable solutions. However, this requires both the  total grain-surface cross-section and elemental S abundance both to have values that are far larger than those inferred for dense molecular clouds (see Table 1). The values adopted here exclude the possibility of  ion-grain neutralization affecting the occurrence of  bistable solutions.

\subsubsection{Desorption}   \label{sec:des}   
 
We adopt simple thermal desorption as the mechanism for returning volatile atoms and molecules to the gas. A neutral molecule $i$ thermally desorbs at a rate
\begin{equation} \label{eqn:des}
\xi_i ~=~\nu_i {\rm exp}( - {{E_{ i}}\over {kT_{\rm gr}}
 })
\end{equation} 
where $T_{\rm gr}$ is the grain temperature, $\nu _i = 10^{12}$~s$^{-1}$ is the vibrational frequency of
 $i$ in a surface binding site and $E_{ i}$ is its binding energy for physisorption.
With the $E_{ i}$ defined, the exponential dependence on $T_{\rm gr}$ means that it largely determines the gas-dust   bifurcation parameters for each autocatalyst-dimer pair, $\rm X-X_2$, i.e.  $\lambda_k  / \xi_k, k=\rm O, O_2$.  

Thermal desorption is actually the most restrictive choice for this mechanism. ISM thermal physics in dark clouds predicts $T \approx T_{\rm gr}$, thus affecting  $\lambda_i$  through the $T^{1/2}$ dependence,  nonlinearly coupling the desorption and accretion rates, and the bifurcations.  This coupling does not occur for nonthermal desorption rates, $\xi_{_{\rm NT}}$. 
We briefly discuss other possible desorption mechanisms in $\S \ref{sec:discuss}$.

\subsubsection{Grain-surface reactions}   \label{sec:gsreac}   

    Assuming that all accreted reactive species are instantaneously hydrogenated (e.g. C, CH, CH$_2$, and CH$_3$ become CH$_4$;  N, NH, and NH$_2$ become NH$_3$;  O and OH become H$_2$O; S and HS become H$_2$S, (e.g. Brown \& Charnley 1990) means that no desorption can occur and that there will be no nonlinear feedback.
   As the gas-dust exchange of particles is accounted for  by equations  (\ref{eqnydot}) and  (\ref{eqngdot}), the simplest approach is to neglect  surface reactions. Our neglect of  surface hydrogenation in particular  will be justified {\it a posteriori} as we find that nonlinear kinetic effects become important  when   $T_{\rm gr} \approx 15$K and so H atom thermal desorption can be expected to suppress surface hydrogenation reactions (E$ _{ H}$ $\approx$ 600K).
We return to this issue in $\S \ref{sec:discuss}$.

\section{Chemical model}   \label{sec:model}
Table \ref{tab:table1}  lists the physical parameters of the reference dense cloud model. 
  We employ the same gas-phase  chemical reaction network as in Paper II. 
 The reactions are taken from the UMIST 2012 database (\citet{2013A&A...550A..36M}) and consists for 2001 reactions  involving 
 136 chemical species containing   H, He, C, O, N, S, Na, and Si atoms.   Only photoprocesses resulting from  cosmic-ray-induced photons are considered (\citet{1983ApJ...267..603P}). 
   We only consider a single value for the H$_3^+$ recombination rate $\rm \alpha_3$ (cf. Paper I). 
The chemical evolution is solved  by integrating the  system of differential equations in $\S$ \ref{sec:bistable}  with all species initially present as atoms, except for H$_2$. 
    \begin{table}
        \caption{Gas-grain Chemical Model Parameters  }
        \vspace{2ex} 
        \begin{center}
        \begin{tabular}{lll} 
        \hline
        \hline
        Hydrogen number density $(\rm cm^{-3}) $& $n_ {\rm H}$ & $1.0\times 10^{4}$; $1.0\times 10^{5}$ \\[0.1cm]
        Cosmic ray ionization rate ($\rm s^{-1} $) & $\zeta_{_ {\rm 0}}$ & $1.3\times 10^{-17}$ \\[0.1cm]
        Dust temperature \rm (K) & $T_{\rm gr}$ & $\rm see ~ text$\\  [0.2cm] 
             Visual extinction \rm (mag.) & $A_{\rm V}$ & 10  \\[0.2cm]
           H$_3^+$ electron recombination rate ($\rm cm^{3}~s^{-1})$$^\dagger$& $\alpha_{_{\rm 3}}$ & $6.7\times 10^{-8}   T_3^{-0.52}$       \\[0.1cm]
         Total dust cross-section ($\rm cm^{2}$) & $D^{^{\rm 0}}  _{\rm gr}$ & $10^{-22} \pi$  \\[0.1cm]
        Gas temperature \rm (K) & $T$ &  $T_{\rm gr}$   \\[0.4cm]

           
       Total  elemental abundances:  
       & $ Y^{\rm T}_{_ {\rm   He }}   $    &  0.14\\   [0.1cm]    
        &  $ Y^{\rm T}_{_ {\rm O}} $   & $1.6\times 10^{-4}$  \\ [0.1cm]
 &  $ Y^{\rm T}_{_ {\rm C  }}   $   &$6.0\times 10^{-5}$ \\ [0.1cm]
 &  $ Y^{\rm T}_{_ {\rm N  }}   $   & $2.0\times 10^{-5}$  \\ [0.1cm]    
    & $ Y^{\rm T}_{_ {\rm  S }}   $   & $2.0\times 10^{-8}$\\[0.1cm] 
    & $ Y^{\rm T}_{_ {\rm  Si }}   $   & $ 0 $\\ 
   &$ Y^{\rm T}_{_ {\rm Na  }}  $       & $2.0\times 10^{-9}$\\ [0.1cm]     
      
                         Surface binding energies (K)$^\ddagger$:  \\[0.1cm]
                         
        &  $ E_{_ {\rm O}} $     &$ \rm 800K$  \\ [0.1cm] 
        &  $ E_{_ {\rm N}} $   & $ \rm 800K$  \\ [0.1cm] 
        &  $ E_{_ {\rm C}} $    & $ \rm 800K$ \\ [0.1cm] 
        &  $ E_{_ {\rm S}} $    & $ \rm 1100K$ \\  [0.1cm]
         &   $ E_{_ {\rm O_2}} $    &  $ \rm 1000K$  \\  [0.1cm]
        & $ E_{_ {\rm N_2}} $   & $ \rm 790K$   \\  [0.1cm]
        &   $ E_{_ {\rm CO}} $    &  $ \rm 1150K$ \\  [0.1cm]

                 \hline 
        \end{tabular}
        \end{center}
        $^\dagger T_3$=$T$/300K\\
        $^\ddagger$ For the most volatile species; \citet{2013A&A...550A..36M}.
         \label{tab:table1}
        \end{table}

\subsection{Surface binding energies}   \label{sec:surfbind} 

For consistency, we  have adopted the binding energies,  $ E_{_ {\rm i}} $,  listed  in the UMIST database  (\citet{2013A&A...550A..36M}) for this initial study.   Alternative compilations exist (e.g. \citet{2017ApJ...844...71P}) but we take the UMIST values as a reference for  comparison with future studies.   The binding energies of helium atoms and  H$_2$  molecules are so  small  that they do not stick to grains even at at the low temperatures of our models.  
For atomic C, O and N we have taken the single value  $ E_{_ {\rm i}} =800$K, based on simple polarizability arguments (\citet{1987ASSL..134..397T}), and so there is no difference in the  desorption rates of these species that could complicate the analysis. 

The UMIST $ E_{_ {\rm i}} $ values are consistent with recent calculations and experiments for O$_2$, CO, N and N$_2$   (\citet{2017ApJ...844...71P, 2017MolAs...6...22W,2020ApJ...904...11F,2020MNRAS.499.1373M}) but recent work indicates differences  in the binding energies of C and O atoms. The influence of O atom binding energies on the models are discussed in $\S$ \ref{sec:discuss}. 
For C atoms, calculations by \citet{2018ApJ...855...27S} indicate that they bind to water by  chemisorption with $ E_{_ {\rm C}} \sim 16000$K. 
Our results should not be particularly sensitive to the atomic C  thermal desorption rate since the chemical evolution of interest in our models only  begins  at around the gas-grain accretion timescale,   at which point most of the initially abundant atomic C  has already been incorporated into CO (after $\sim \rm few \times 10^5$ years).

\section{Bifurcations and oscillations in interstellar oxygen chemistry} \label{sec:heho}
  
We first consider a simple reduced model,  containing only hydrogen, helium and oxygen, that was previously studied in Papers I \& II.  This model  has two possible autocatalytic pathways in the O-O$_2$ chemistry and is useful for understanding the network kinetics before considering larger, more realistic, chemical models.
In the first instance,   for this model and those considered later, we are primarily interested in the nonlinear dynamical evolution of the system of ODEs.  We therefore integrate over time-scales ($\sim$10$^7$-10$^8$ yrs) longer than  the estimated lifetime of molecular clouds (\citet{2012ARA&A..50..531K}) or the time-scale to reach a gas-phase chemical steady-state (in the absence of accretion onto dust) so that the evolution of any instabilities can be followed in the long term.
    
\begin{figure}
\includegraphics[width=0.5\textwidth]{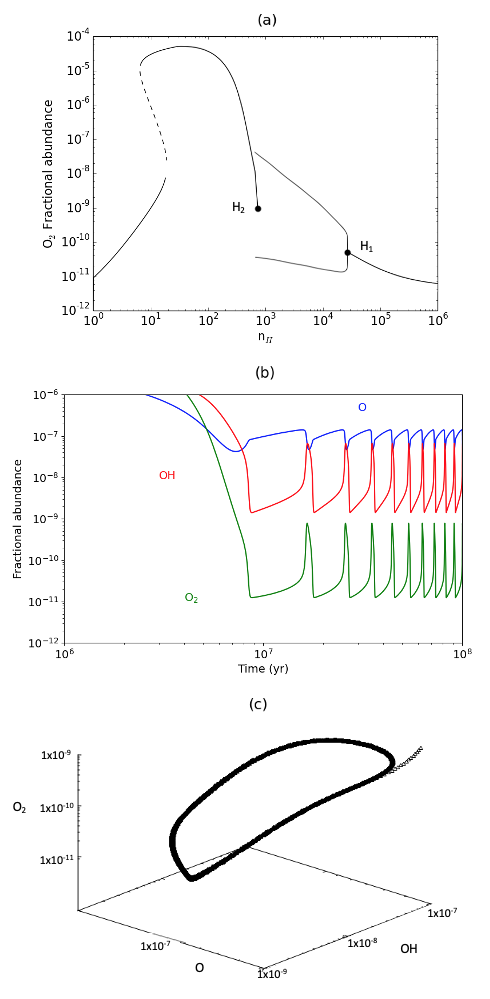}   
\caption{(a) Bifurcation diagram for the O$_2$ abundance in the reduced model for $T_{\rm gr} =14.70$K; $H_1$ and $H_2$ denote the Hopf bifurcation points;  
(b) Time series for $ y_{_ {\rm O}}(t)$, $ y_{_ {\rm OH}}(t)$ and $ y_{_ {\rm O_2}}(t)$ at $ n_ {\rm H} = 1\times 10^{4}$ $\rm cm^{-3}$ in the bifurcation diagram; 
(c) Phase space evolution of the time series of (b); 
 }
\label{fig:fig1}
\end{figure}

  For the model parameters of Table 1, we compute the bifurcation diagram as a function of   $n_ {\rm H}$ for different values of   $T_{\rm gr}$.  We find that  that oscillating solutions can exist  in a limited range $T_{\rm gr} \approx 14.5-15$K.  Figure \ref{fig:fig1}(a) shows the bifurcation diagram for the case of $T_{\rm gr} =14.70$K.
  The bistable stationary states at low  $n_ {\rm H}$  involve  a {\it fold} bifurcation as found in earlier studies.   It  is due entirely to gas-phase chemistry and is set by $\zeta/n_ {\rm H}$  (\citet{1998A&A...334.1047L}) and $ Y^{\rm T}_{_ {\rm O}} $; depletion of species onto dust is negligible and so the abundance of O nuclei in the gas, $ Y_{_ {\rm O}} $, approximately equals $ Y^{\rm T}_{_ {\rm O}} $. From Paper II, neglecting dust, as  $ Y^{\rm T}_{_ {\rm O}} $ is reduced this bifurcation moves to higher densities.
  Inclusion of the gas-dust exchange of species leads to $ Y_{_ {\rm O}} $ being  set by accretion-desorption of oxygen atoms and molecules and the emergence of  two {\it Hopf} bifurcations, where the system transitions from stationary to periodic solutions. 
  The Hopf points $H_1$ and $H_2$, cover the interval    $  7\times 10^{2} \lesssim n_ {\rm H} \lesssim  2\times 10^{4}$ $\rm cm^{-3}$, and  $ Y_{_ {\rm O}} \ll Y^{\rm T}_{_ {\rm O}} $.
  
 We were unable to fully characterize the  Hopf bifurcations using Newton-Raphson iteration due to problems with branching and continuation to obtain the unstable solutions.  The periodic solutions shown are those  obtained with the ODE solver when carried out to  10$^8$ years. This approach was based on the premise that if stable periodic (limit cycle) oscillations emerge they will still be present on this time-scale and the amplitude maxima and minima for  each species, $ y_{_ {\rm max}}$ and $ y_{_ {\rm min}}$, should be constant; the solutions presented thus plot these ($ y_{_ {\rm max}},  y_{_ {\rm min}}$) pairs at each $n_ {\rm H}$.

   The Hopf bifurcations in  Figure \ref{fig:fig1}(a) can be classified as follows (\citet{seydel2009practical}).
  $H_1$ is a  supercritical Hopf bifurcation. Periodic solutions are present in the immediate vicinity of $H_1$  where there is a {\it soft } transition from the stationary state.    $H_1$ has a pitchfork structure with an unstable stationary state connecting it to $H_2$ (not shown).           
 $H_2$ is a subcritical Hopf bifurcation where, in this case, the periodic solutions occur after a jump from the  stationary state, i.e. a {\it hard } transition. $H_2$ is connected to the two adjacent stable  periodic solutions by two unstable periodic solutions (not shown).  
 
Figure \ref{fig:fig1}(b) shows the time series for the oscillatory solutions for $ y_{_ {\rm O}}(t)$, $ y_{_ {\rm OH}}(t)$ and $ y_{_ {\rm O_2}}(t)$ at $ n_ {\rm H} = 1\times 10^{4}$ $\rm cm^{-3}$. These are limit cycle oscillations, as confirmed in the phase plane evolution shown in Figure \ref{fig:fig1}(c).    

\section{Relaxation oscillations in molecular clouds}\label{sec:chemosc}

In Papers I \&  II we demonstrated that the simple model of the preceding section is bistable for all values of $ Y^{\rm T}_{_ {\rm O}} $   and that these solutions can persist in more realistic molecular cloud chemical models. We now show that this is also the case for chemical  oscillations. 
   Figure \ref{fig:fig2} shows the chemical evolution for the reference dense cloud model at $ n_ {\rm H} = 1\times 10^{4}$ $\rm cm^{-3}$ and two values of  $D_{\rm gr}$. We find that slightly higher values of $T_{\rm gr} $ are required for oscillations to appear than in the pure oxygen chemistry because the presence of other elements (C,N,S) can shift the range of $\zeta/n_ {\rm H}$ for bistability to higher $ n_ {\rm H} $ (Paper I). 
    
\subsection{Relaxation oscillations at n$_H$ = 1$\times$10$^4$ cm$^{-3}$}\label{sec:chemosc1}

 Figure \ref{fig:fig2}(a) shows that {\it nonlinear relaxation oscillations} are present for $T_{\rm gr}=15.26$K. The time series of the abundance oscillations are different from those found in  $\S$ \ref{sec:heho} and are generally characterised by a `ringdown' of decaying amplitude and increasing period. 
Gaseous O, O$_2$, N and N$_2$  are regulated by thermal desorption and the amplitude of the oscillations corresponds to variations of  $\sim 10^2-10^3$    in the OH, O$_2$ and H$_2$O abundances. NH$_3$ can oscillate in abundance by a factor of $\sim 10^5$.
The nitrogen chemistry is coupled to that of oxygen through the reaction sequence that catalyses N$_2$ formation in molecular clouds 
    \begin{equation}
  \rm{ N 
  \buildrel OH \over \longrightarrow NO
  \buildrel N  \over \longrightarrow N_2 ~+~ O.}
  \label{ncycle}
\end{equation}
An important point is that the nitrogen chemistry does not directly affect the oxygen chemistry: no oxygen is directly lost in the sequence (\ref{ncycle}) and the reaction 
\begin{equation}
{\rm N ~+~ O_2 ~  \longrightarrow  ~ NO  ~+~ O}\label{n+o2}
\end{equation}
has a negligible rate at low temperatures (e.g. \citet{2013A&A...550A..36M}).
 As noted in $\S$ \ref{sec:des},  CO formation through 
    \begin{equation}
{\rm C ~+~ OH ~  \longrightarrow  ~ CO  ~+~ H}\label{c+oh}
\end{equation}
\begin{equation}
{\rm C ~+~ O_2 ~  \longrightarrow  ~  CO  ~+~ O}\label{c+o2}
\end{equation}
substantially depletes atomic carbon from the gas within an accretion timescale. This has the effect that, at   $T_{\rm gr} \approx 15$K, almost all the carbon in CO is frozen out.
 Other carbon-bearing molecules  also freeze out, as do the  sulfur-bearing species. Although the adopted binding energy of atomic S is comparable to that of O$_2$ (Table 1), it is converted in gas-phase reactions to species with higher binding energies: SO, SO$_2$ and CS, which do freeze out efficiently.

Our conservative choice of dust parameters (Table 1) and the gas density leads to the nonlinear oscillations appearing, as in the H-He-O model,  on timescales ($\gtrsim$ $10^7$ years)  that exceed estimated molecular cloud lifetimes. We note that some estimates of Milky Way GMC lifetimes ($\sim$1.5-4$\times$ 10$^7$ years) are in range where chemical oscillations could arise (\citet{2011ApJ...729..133M}). Figure \ref{fig:fig2}(b) shows the chemical evolution in a model with $D_{\rm gr}=5D^{^{\rm 0}}_{\rm gr}$ and   $T_{\rm gr} = 15.46K$.  
With a slight increase of  $T_{\rm gr}$ and a  larger dust cross-section,   the  oscillations can be made to begin at earlier times ($\gtrsim $4 $\times 10^6$ years),  but with a lower maximum amplitude than in 2(a).
Solutions with larger amplitude oscillations  should also exist but are difficult to find without a bifurcation diagram for the system; we return to this point in $\S$ \ref{sec:discuss}.
 
 
 The presence of the  relaxation oscillations is a characteristic of a nonlinear dynamical system that has undergone Hopf bifurcation but in which some of the  bifurcation parameters vary slowly in time (e.g. \citet{seydel2009practical}).
  In this model the combination of bifurcation parameters  [$\zeta/n_ {\rm H} ;  \lambda_k  / \xi_k, k=\rm O, O_2$] are fixed (at given $T_{\rm gr}$)  and, to first order,   determine the appearance of the Hopf bifurcation (recall that these also initially fix $ Y_{_ {\rm O}}$). 
  However, chemical reactions convert O, N and C  nuclei in the gas into molecules that can be permanently removed  by accretion, depending on  $T_{\rm gr}$.  This means that the  gaseous elemental abundances $Y_{_ {\rm O}}(t)$, $ Y_{_ {\rm N}}(t)$ and $ Y_{_ {\rm C}}(t)$ undergo a slow quasi-stationary drift in time; even very small changes  can have dramatic effects on the evolution. 
  Specifically,   $ Y_{_ {\rm N}}(t)$ declines due to loss of N nuclei in NH$_3$,  NH$_2$, NH and NO;  this affects  the oxygen chemistry though the sequence (\ref{ncycle}) and by the direct removal of O nuclei by NO accretion. 
 The slow  conversion of the trace amounts of atomic C to CO in reactions (\ref{c+oh}) and (\ref{c+o2}), followed by  CO accretion, also influences  $ Y_{_ {\rm O}}(t)$ and $ Y_{_ {\rm C}}(t)$.

\subsection{Relaxation oscillations at n$_H$ = 1$\times$10$^5$ cm$^{-3}$}\label{sec:chemosc2}
 
These chemical models are similar in composition to observed  cloud cores in which gaseous CO has a low abundance, disappearing at the highest densities  while those of N$_2$ ($\rm N_2H^+$) and NH$_3$ can still be  present (e.g. \citet{2007ARA&A..45..339B,2012ApJ...757L..11W,2019MNRAS.487.1269S}).  Such  CO-depleted cores typically have  $\sim 10-100$ times greater densities and so  we next consider a higher density model. 

Adopting  $ n_ {\rm H} = 1\times 10^{5}$ $\rm cm^{-3}$, we also set $\zeta = 10\zeta_{_ {\rm 0}}$   to maintain the same $\zeta/n_ {\rm H}$ ratio so that a gas-phase bistable solution can approximately be recovered.  As before, slightly higher  $T_{\rm gr}$ values are required for oscillations to occur. Figure \ref{fig:fig3} shows the time series for the long-term evolution of selected species as a function of $T_{\rm gr}$ and that relaxation oscillations now begin after one accretion time ($\gtrsim  10^6$ years), a time-scale that is relevant for comparison with molecular cloud chemistry.

\subsection{Initial conditions}
The assumption of completely atomic initial conditions for elemental C, O, N, S neglects the fact that dense cloud gas is presented by a diffuse/translucent cloud phase and so initial molecular abundances should realistically be non-zero (cf. \citet{2006ARA&A..44..367S}). \citet{2020A&A...643A.121R} have demonstrated that the choice of initial conditions, specifically the CO/C$^+$ or CO/C ratios, can lead to different 'early-time' chemistry which affects the subsequent evolution of the purely gas-phase oscillations. 

\begin{figure*}
\includegraphics[width=0.9\textwidth]{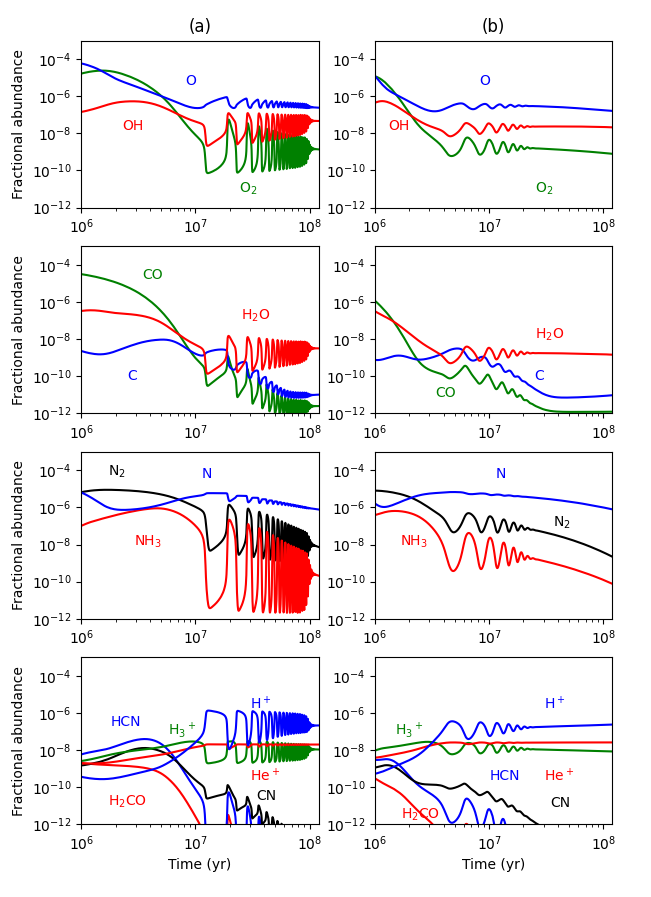}   
\caption{Gas-grain chemical evolution in a dense molecular cloud  for $\zeta = \zeta_{_ {\rm 0}}$ and  $ n_ {\rm H} = 1\times 10^{4}$ $\rm cm^{-3}$: 
(a) Left-hand panels, $D_{\rm gr}=D^{^{\rm 0}}_{\rm gr}$, $T_{\rm gr} = 15.26K$; (b) Right-hand panels $D_{\rm gr}=5D^{^{\rm 0}}_{\rm gr}$,   $T_{\rm gr} = 15.46$K.}
\label{fig:fig2}
\end{figure*}

\begin{figure*}
\includegraphics[width=0.9\textwidth]{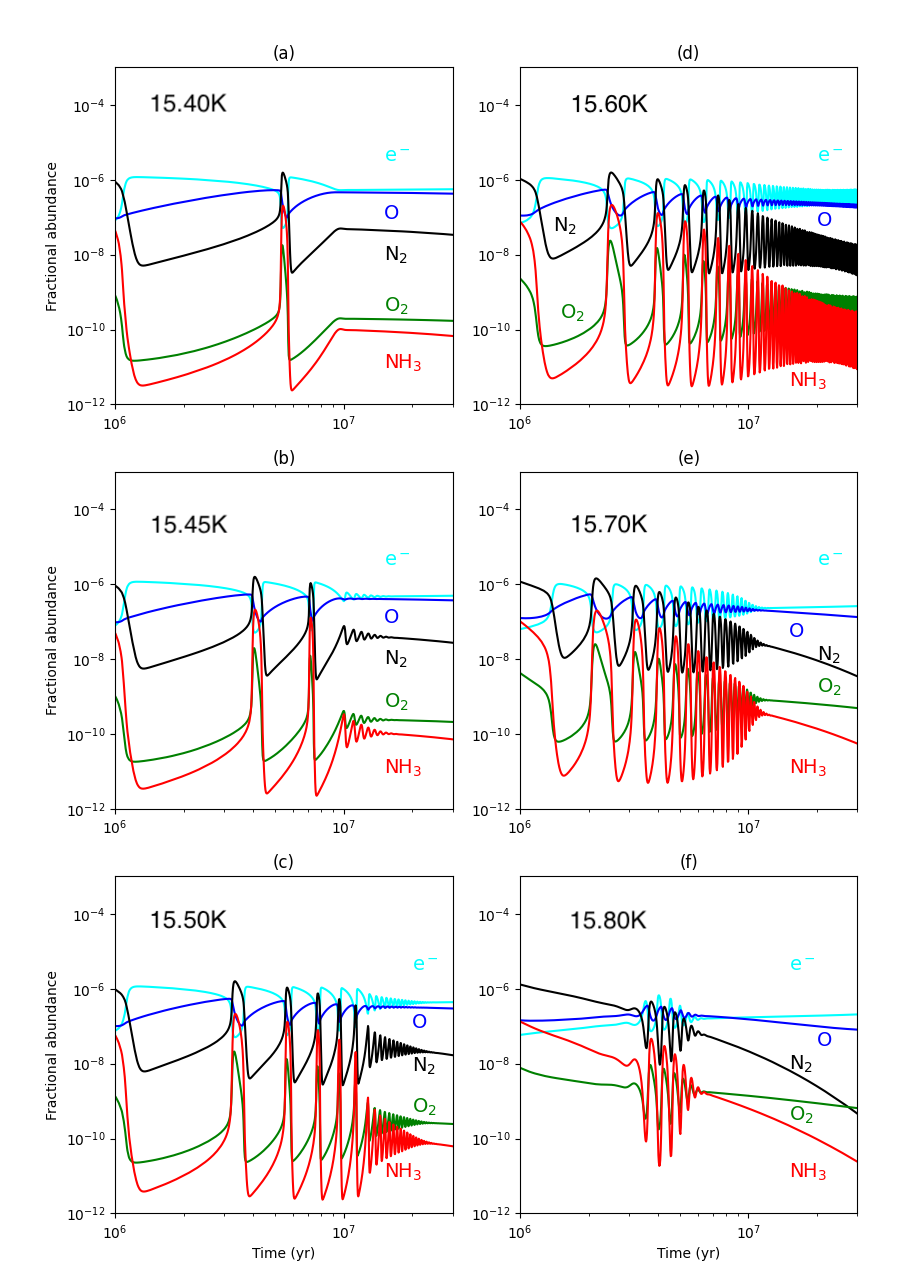}   
\caption{Gas-grain chemical evolution of selected species as a function of the bifurcation parameter $T_{\rm gr} $.   This model has  $\zeta = 10\zeta_{_ {\rm 0}}$, $ n_ {\rm H} = 1\times 10^{5}$ $\rm cm^{-3}$, $D_{\rm gr}=D^{^{\rm 0}}_{\rm gr}$  and  the $T_{\rm gr} $ are given in the upper left-hand corner of each panel.}
\label{fig:fig3}
\end{figure*}

As the oscillations reported here are controlled by the gas-dust exchange of material; they occur on time-scales much longer than that of 'early-time' chemistry, and so are likely less sensitive to the adopted initial conditions. We computed dense cloud models in which initial conditions were modified such that almost all carbon is present as C and CO; such that CO/C = 10 (\citet{2010ApJ...708..334B}), with the remaining traced molecules having abundances appropriate for the diffuse ISM (\citet{2006ARA&A..44..367S}, Figure \ref{fig:fig2}). We find that these initial conditions do not affect the onset or characteristics of the gas-grain oscillations found here. 

 \section{Discussion}\label{sec:discuss}
  
The gas-grain models considered here, evolve to produce the conditions of  CO-depleted dense cores (e.g. \citet{2007ARA&A..45..339B}).  
When the mechanism mediating the gas-grain exchange of molecules is  thermal desorption, grain  temperatures of  $T_{\rm gr} \gtrsim$ 20K would be required for the models to maintain significant CO abundances at later times.  The C and C$^+$ released from CO could then influence the O$_2$-O autocatalysis and the nonlinear solutions found here ($\S$ \ref{sec:chemosc}, Paper I). 
Furthermore, alternatives to the  UMIST $ E_{_ {\rm i}} $ values of Table 1 exist\footnote{See \citet{2022ESC.....6..597M} for a recent review of the physics of thermal desorption relevant to interstellar dust.}      experimental measurements of $ E_{_ {\rm i}} $ for  O atom binding  on water ice  lie in the range $1440-1660$K (\citet{2015ApJ...801..120H,2022ESC.....6..597M}).        
Thus, to produce oscillations in these models via O$_2$-O desorption requires higher dust temperatures to match the desorption rates considered here; $ \xi{_ {\rm O}}(T_{\rm gr}) $ and $ \xi{_ {\rm O_2}}(T_{\rm gr}) $ at 15K requires $T_{\rm gr} \approx 27-31$K. 
However, while  $T_{\rm gr}$ in this range will also  desorb CO and other molecules, making the chemistry similar  to that of dense clouds,   the  $\zeta/n_ {\rm H}$ range  for bistability in the gas will be different  due to the presence of the additional elements, particularly C and S (Paper I) and so a search of the  parameter space for Hopf bifurcations would be required.

In fact, other autocatalytic cycles and bistable solutions are present in interstellar chemistry (Paper II) and these could also be made to undergo Hopf bifurcation.  Atomic nitrogen can act as an autocatalyst and bistable states appear at densities of  $\rm  \sim 10^5$ cm$^{-3}$ and higher,  for all values of $ Y^{\rm T}_{_ {\rm N}} $. 
Although we did not consider grain-surface chemistry explicitly,  we note that estimates of the binding energies of CH$_n$ with n=1,3;   species are low enough  (\citet{2017ApJ...844...71P,2017MolAs...6...22W}) that they could be desorbed, along with CH$_4$, either thermally or by reactive desorption (e.g. \citet{1990MNRAS.244..432B}).  As gas-phase carbon chemistry has an autocatalytic cycle, nonlinear oscillations could then be generated by gas-dust exchange of the  methylidyne radical  autocatalyst (CH).       
      
%
  
Other desorption mechanisms are possible, including non-thermal desorption and reactive desorption (e.g. \citet{1985A&A...144..147L,2016A&A...585A..24M,2018ApJ...853..102C,2021ApJ...922..126S,2021A&A...652A..63W}).  
As long as the rates of any alternative desorption mechanism returns O and O$_2$ at the same rates as  those considered here  oscillations should still occur,  i.e. for non-thermal desorption $ \xi_{_ {\rm NT}} \sim  \xi{_ {\rm O}} ~ (T_{\rm gr}) \approx 15$K). 
Our assumption that  $T=T_{\rm gr}$ is consistent with  thermal balance  in dense clouds but is not necessarily applicable when non-thermal desorption is considered.   However, we find that  models  in which this condition is relaxed  ($T \neq T_{\rm gr}$)  produce similar dynamical solutions.  
Reactive desorption of grain-surface reaction products could, in principle,  still produce oscillations if the sum of the their  desorption rates,  and the rate of their subsequent  breakdown in gas-phase reactions to an X$_2$-X pair,  matches the thermal desorption rates found here.

In fact, in a parameter-space (grid model) study of dark cloud chemistry, in which reactive desorption was employed as the sole desorption mechanism, \citet{2018ApJ...868...41M} claimed to have found oscillatory solutions for some species, specifically for the abundances of long carbon-chain molecules. However, we can discount the presence of limit cycles or relaxation oscillations in the model of Maffucci et al. since, if present, all species should oscillate in abundance (cf. \citet{2020A&A...643A.121R}) 

 In a recent grid model study, \citet{2022A&A...662A..39E} found one model in which that oscillations could develop. This model had very large cosmic ray ionization rate ($\zeta = 10^{-13} \rm s^{-1}$) with molecules desorbed through photodesorption by cosmic ray-induced photons. These oscillations began at $\sim 10^{3}$ years and ended at $\sim 10^{3}$ years once molecules had been significantly destroyed. This early onset time-scale indicates that the gas-grain interaction was not involved. Nevertheless, this points to another astrophysical environment where exotic kinetics could occur.

Finally,  the  above brings us to the important point that  searching  for oscillations without bifurcation diagrams (i.e. employing ODEs or scaling from known solutions) is risky.  Their appearance depends sensitively on the bifurcation parameters,  $\zeta/n_ {\rm H}$ and   $\lambda_k  / \xi_k$, which are are coupled through  $n_ {\rm H}$  and $T$ (when $T = T_{\rm gr}$ is assumed),  and the relative elemental abundances.

To find Hopf bifurcations and oscillations would first require a bifurcation analysis of the full CHONS  gas-phase chemistry to locate the range of $\zeta/n_ {\rm H}$ and  elemental abundances to produce bistable solutions  (Dufour \& Charnley 2023, in preparation).  This can then be followed by   mathematical analysis of the stability of the solutions for the gas-grain chemistry (\citet{1996OxPres...314...G}). Analysis of this model from the perspective of the non-linear dynamics is beyond the scope of this article and will be considered elsewhere.

\section{Conclusions}\label{sec:conc}

Autocatalysis in interstellar gas-phase chemistry can lead to bistability which, when   coupled with the gas-grain exchange of key species, can undergo Hopf bifurcation and lead to the appearance  of limit cycles and chemical relaxation oscillations.   
The nonlinear oscillations in these models  occur on time-scales relevant for comparison with molecular cloud composition, particularly the first few periods. 
 
These results complement the oscillations recently presented by \citet{2020A&A...643A.121R}, except that, in realistic molecular clouds, gas-grain oscillations are not permanent limit cycles (see $\S$ \ref{sec:theory}, Figure \ref{fig:fig1}). Both these studies indicate the early-time/late-time dichotomy of molecular cloud modeling can be broken,  and  predict dramatically different chemical evolution from that found heretofore. 
The chemical processes that determine the oscillatory solutions  are  central to almost all chemical models of astrophysical sources, including interstellar clouds and protoplanetary disks.  Oscillations are  therefore fundamental outcomes in astrochemical kinetics, indicating that a wealth of non-linear dynamical phenomena can be present in cold molecular clouds. 
          
This research was supported by the NASA Planetary Science Division Internal Scientist Funding Program through the Fundamental Laboratory Research work package (FLaRe) and by the NASA Goddard Science Innovation Fund. 

\section{Data availability}
The data underlying this article are available in the article and in its online supplementary material.  

\bibliographystyle{mnras}
\bibliography{biblio}

\label{lastpage} 
\end{document}